\documentstyle[aps,prl,multicol,epsf]{revtex}

\def\be{\begin{equation}}
\def\ee{\end{equation}}
\def\bea{\begin{eqnarray}}
\def\eea{\end{eqnarray}}
\newcommand{\corr}[1]{\langle #1\rangle}
\def\br{{\bf r}}
\def\eps{\epsilon}
\newcommand{\Tr}{\mathop{\rm Tr}}
\newcommand{\pp}{{\textstyle\frac\pi2}}

\begin{document}

\title{Quantum superconductor-metal transition in a proximity array}
\author{M.~V.~Feigel'man$^1$, A.~I.~Larkin$^{1,2}$ and M.~A.~Skvortsov$^1$}
\address{$^1$L. D. Landau Institute for Theoretical Physics, 
  Moscow 117940, Russia}
\address{$^2$Theoretical Physics Institute, University of Minnesota,
  Minneapolis, MN 55455, USA}

\date{\today}

\maketitle

\begin{abstract}
A theory of the zero-temperature superconductor-metal transition is
developed for an array of superconductive islands (of size $d$)
coupled via a disordered two-dimensional conductor with the
dimensionless conductance $g = \hbar/e^2R_\Box \gg 1$. At $T=0$
macroscopically superconductive state of the array with lattice
spacing $b \gg d$ is destroyed at
$g < g_c \approx 0.1 \ln^2(b/d)$.
At high temperatures the normal-state resistance between neighboring
islands at $b = b_c$ is much smaller than $R_Q = h/4e^2$.
\end{abstract}

\draft

\begin{multicols}{2}

In two-dimensional (2D) systems, qualitative arguments
based on duality between Cooper pairs and vortices lead to the
prediction~\cite{mpafish} that the superconductor-insulator
transition happens at the universal quantum value $ R_Q \equiv h/4e^2$
of the
resistance per square $R_\Box $. Although a number
of experiments (cf.~\cite{Goldm} and references therein) seem to be in agreement
with this prediction,  other data demonstrate strong deviations
from it~\cite{delft,gantm,valles,kapit}.
Phenomenological picture of duality is not able to
predict the system's parameters (e.~g., the
value of the normal-state resistance) leading to the
quantum critical point: a microscopic theory is needed to find it.
Competition between Josephson coupling $E_J$ and charging energy
$E_C$ is known~\cite{efetov,euro} to be the driving mechanism of
zero-temperature phase transitions between the superconductive and
insulating states in artificial arrays~\cite{chalmers,delft},
films~\cite{2Dgrains,Goldman}
and bulk materials~\cite{3Dgrains}.
In such systems there are no free electrons at very low temperatures due
to Cooper pairing, but pairs may become localized due to Coulomb
repulsion. This is the ``bosonic" mechanism of superconductivity
suppression.

Homogeneously disordered superconductive
films \cite{old,gantm,valles,kapit} present another group of systems
where quantum fluctuations lead to destruction of superconductivity.
The theory of $T_c$ suppression in such films was developed in
Refs.~\cite{finkel}. The qualitative idea behind this theory is that
disorder-enhanced Coulomb repulsion leads to the decrease of Cooper
attraction and thus to the decrease of $T_c$. The superconductive
transition temperature vanishes~\cite{finkel,finkel3} when the
dimensionless film conductance $g = \hbar/e^2R_\Box$ decreases down
to $g_{\rm Fin} = (2\pi)^{-2}\ln^2(1/T_{c0}\tau_{\rm tr})$, where
$T_{c0}$ is the BCS transition temperature and $\tau_{\rm tr}$ is the
elastic scattering time.
This second (``fermionic") mechanism of superconductivity suppression
is clearly different
from the first one~\cite{efetov} since its basic feature lies
in disappearance of Cooper pairs altogether. Experimental data
supporting the
fermionic mechanism are reviewed in Ref.~\cite{finkel3}. A drawback
of this theory is that it
neglects
quantum fluctuations of
the bosonic field (i.~e., it can be considered as a kind of the BCS
theory with the renormalized attraction constant).
For a phenomenological comparison of the bosonic and fermionic
mechanisms, cf.\ Ref.~\cite{Hamburg99}.

In this paper we study a model for quantum breakdown of
superconductivity, which lies in between the two limiting cases
discussed above.
We consider an array of small superconductive (SC) islands (of radius
$d$ each) in contact with a thin film of dirty normal (N) conductor
with the dimensionless conductance $g \gg 1$. 
%% The concentration of islands is $b^{-2}$, with $b \gg d$.
The distance between neighboring islands is $b \gg d$
(more precisely, $b^{-2}$ will be the concentration of islands).
Resistance $R_T$ of the interface
between each island and the film is low: $G_T = \hbar/e^2R_T \gg 1$.
Islands are thick enough, to prevent suppression of superconductivity
inside them. The corresponding condition for the superconductive gap reads
$\Delta_{SC} \gg G_T/\nu V_i$, where $V_i$ is the island's volume and
$\nu$ is the density of states. The above inequality ensures that
the lifetime of electrons in the SC island
(which is finite due to the tunneling across the interface) is much
longer than $\hbar/\Delta_{SC}$.
We assume also that $G_T^2 \gg 4\pi g$; the meaning of this condition will
be explained below.
We will show that
macroscopic
superconductivity in such a system at $T=0$ becomes unstable with
respect to quantum fluctuations
at $g$ less than
\be
  g_c = {\cal G}_c\left(\frac{1}{\pi}\ln\frac{b}{\tilde{d}}\right)^2,
\label{main}\
\ee
where
${\cal G}_c \sim 1$ will be determined below, and $\tilde{d} \sim d$.
Eq.~(\ref{main})
presents our main result (obtained within logarithmic accuracy),
which shows that
the critical sheet resistance $R_{\Box c} = \hbar/e^2g_c$ is much
less than the quantum resistance $R_Q$, provided $\ln(b/d) \geq 3$.
Moreover, the same is valid for the normal-state resistance between
neighboring islands
$R_n = (R_{\Box c}/\pi)\ln(b/d) \sim 3R_Q/\ln(b/d)$.
This result is at odds with
usual arguments based on the model of resistively shunted
Josephson junctions~\cite{schmid}: in that model superconductive behavior
of a single junction is preserved at $T=0$ as long as
it has resistance  $R < R_Q$. The physical reason for this discrepancy
is that we account for the discrete
nature of charge transport between the SC islands, which was neglected
in the Caldeira-Leggett model adopted in~\cite{schmid}.

We will follow an idea presented in Ref.~\cite{FL}, where a
simplified version of the considered model was analyzed (cf.\
also~\cite{SZH}). Namely, we make use of the long-range nature of the
Josephson coupling $J_{ij}$ between SC islands due to the proximity
effect in the film, which scales as $J_{ij} \propto r_{ij}^{-2}\exp[-
c(r_{ij}/L_T)]$, where $L_T= \sqrt{\hbar D/T}$ is the thermal
coherence length (cf.\ Refs.~\cite{FL,ALO}), $c \sim1 $, and $D$ is
the diffusion constant in the film.
At low temperatures the interaction radius $L_T$ diverges indicating
that the position of the quantum phase transition can be found in the
mean-field approximation (MFA) analogous to the one developed in
Ref.~\cite{VaksLP}. Within the MFA, macroscopic superconductive
coherence sets in at
\be
\frac1{2\hbar}{\cal J}(T) {\cal C}(T) \geq 1,
\label{transition}
\ee
where
\be
  {\cal J}(T) = \sum_jJ_{ij},
  \qquad
  {\cal C}(T) = \int\nolimits_0^{1/T}d\tau C_0(\tau),
\ee
and $C_0(t) = \langle \cos ({\theta (0)-\theta (t))}\rangle$ is the
single-island autocorrelation function of the order parameter phase.
Thus, to find the superconductor-metal transition line we need to determine
the functions ${\cal J}(T)$ and ${\cal C}(T)$ at low temperatures.
We will discuss the $T=0$ limit mainly, and present the results for
low $T > 0$ at the very end of the paper.

Calculation of ${\cal J}(0)$ is rather straightforward.
The result depends on the relation between the dimensionless
Cooper repulsion constant $\lambda_n$ defined at the energy
scale $\omega_d=D/d^2$ and the ratio $G_T/4\pi g$ which is
a measure of the interface transparency.
At low interface transparencies, $G_T/4\pi g \ll \lambda_n$,
it is possible to consider the lowest order process over $G_T$,
and one obtains~\cite{FL}
\be
  {\cal J}(0) = \frac{G_T^2}{16\nu\lambda_n^2} \frac{1}{b^2\ln(b/d)}.
\ee
This result is similar to the one obtained in~\cite{ALO} for a one-dimensional
S-N-S geometry: in both cases ${\cal J}(0)$ formally
diverges as the Cooper repulsion
constant goes to zero. However, the decrease of $\lambda_n$ at a fixed $G_T$
drives the system out of the low-transparency limit.
At $\lambda_n \leq G_T/4\pi g$  full nonlinear treatment
of the proximity problem is necessary, accounting multiple Andreev
reflections from SC islands.
At the same time, the effect of $\lambda_n$ upon ${\cal J}(0)$
can be neglected in this limit (for 2D geometry studied here).
To find the coupling energy $J(r)$ between two islands
situated at the distance $r \gg d$, one has to solve
the Usadel equation~\cite{Usadel} in the film.
Actually, it is more convenient to calculate the maximum superconducting
current between the islands, $I_c(r) = (2e/\hbar) J(r)$.
The value of the supercurrent can be determined as the integral over
the current density, taken along the straight line which lies in the
middle between two islands, and is perpendicular to the vector $\bf r$
connecting them.
If the distance between the islands is very large, $\ln(r/d) \gg 1$,
anomalous part of the Green function is small along the above-mentioned
line, that allows%
\footnote{Similar calculation can be found in Appendix C of
Ref.~\cite{Qproxy}}
to linearize the Usadel equation at relevant
energies $E \sim \hbar D/r^2$ and find $I_c(r)$.
Thereby one obtains 
% $J(r)= g(\pi^3/4)\hbar D/r^2\ln^2\frac{r}{d}$
$J(r)= (\pi^3/4) g D/r^2 \ln^2(r/d)$
and, finally,
\be
  {\cal J}(0) = \frac{\pi^4}{2}\frac{g D}{b^2 \ln(b/d)} .
  \label{J}
\ee

The key point in the discussion of the $T=0$ transition is to
determine ${\cal C}(T\to 0)$.
We will see that
${\cal C}(0)$ depends exponentially on the film conductance $g$. If
islands do not have ohmic contacts with the film (coupling via
capacitance $C_{j}$ only) then ${\cal C}(0) = \hbar/E_C = \hbar
C_{j}/2e^2$. In our case $\hbar/{\cal C}(0)$
plays the role of an effective charging energy $E_C^*$ of an island
that survives in spite of good conductance around. To make ideas
transparent, we first discuss a simplified model~\cite{FL} with
sufficiently strong Cooper-channel repulsion in the film, $\lambda_n
\gg G_T/4\pi g$. Then dynamics of the
phase $\theta(t)$ of a single SC island can be described by a simple
imaginary-time action,
\be
  S_0[\theta] =
  - \frac{G_A}{8\pi} \int \!\! \int_0^{1/T}
  dt \, dt' \, \frac{\cos[\theta(t)-\theta(t')]}{(t-t')^2}.
\label{Action1}
\ee
Here $G_A = G_T^2/4\pi g\lambda_n$
is the Andreev subgap conductance (normalized to $e^2/\hbar$) in the
limit of weak proximity effect, valid under the condition
$\lambda_n \gg G_T/4\pi g$~\cite{FL,FLS}%
\footnote{
  The result for $G_A$ given in~\cite{FL} is overestimated by the 
  factor of 2 due to a numerical mistake.%
}.
Expression (\ref{Action1}) is
valid at low energies, $\omega \leq \omega_de^{-1/\lambda_n}$,
while at higher energies,
$\omega_de^{-1/\lambda_n} \leq \omega \leq \omega_d$,
one has $G_A(\omega) = (G_T^2/4\pi g)\ln(\omega_d/\omega)$.
Thus, the Andreev conductance is large compared to $4e^2/h$ for all
frequencies below $\omega_d$
if the  condition $G_T^2\gg 4\pi g$ is fulfilled. At lower values of
$G_T$, N-S transport across the interface is suppressed by the usual
Coulomb blockade effect governed by the junction's charging energy
$2e^2/C_j$.  We are not interested in this trivial effect, and will
assume the condition $G_T^2\gg 4\pi g$ to be valid.

For large $G_A$ one can start from
the Gaussian approximation for $S_0[\theta (t)]$. Then the
Fourier-transformed correlator of phase fluctuations
$\langle|\theta_{\omega}|^2\rangle_0 = 4/|\omega|G_A$, and, hence,
$C_0(t) = e^{-\frac{1}{2}\langle (\theta(t)-\theta(0))^2\rangle_0}
\propto t^{-4/\pi G_A}$.
At $G_A > 4/\pi$,
${\cal C}(T\to 0)$ diverges that seems to indicate that at large
$G_A$ superconductivity is always stable at $T=0$, in agreement
with~\cite{schmid}. The crucial point
is to note that the employed Gaussian approximation breaks down at a
finite time scale $t^*$, due to downscale renormalization of $G_A$.
This renormalization is caused
by the periodicity of the action $S_0[\theta]$ as a functional of
$\theta(t)$, that is, in physical terms, by the charge quantization.
This problem is analogous to
the one studied by Kosterlitz~\cite{Kosterlitz}. Translating his
results to the present case, one gets the renormalization group (RG)
equation
$dG_A(\zeta)/d\zeta = -4/\pi$, with
$\zeta = \ln\omega_d t$.
This equation is to be solved with the initial condition
$G_A(0)=G_A$. As a result, at the time scale $t^* \sim \omega_d^{-1}
e^{\pi G_A/4}$ the renormalized Andreev conductance $G_A(t^*)$ decays
down to the value of order unity~\cite{FL}. At longer time scales
$C_0(t)$ decays approximately as $t^{-2}$, so the integral ${\cal
C}(0) \sim t^* \sim \omega_d^{-1} e^{\pi G_A/4} $. Taking into
account that ${\cal J} \sim b^{-2}$, and using
Eq.~(\ref{transition}), one obtains~\cite{FL} the critical distance
between islands $b_c \sim d e^{\pi G_A/8}$.

However, this result is valid
under the condition $\lambda_n \gg G_T/4\pi g$
which is difficult to
realize simultaneously with the inequality $G_T^2 \gg 4\pi g$
needed to avoid trivial Coulomb blockade effects. Indeed, at
energies $E\ll \hbar/\tau_{\rm tr}$, Cooper interaction constant
$\lambda(E)$ is determined by the RG equation~\cite{finkel} which we
present in a simplified form~\cite{FLS} valid for
$\ln(\hbar/E\tau_{tr}) \ll g$ when renormalization of $g$ can be
neglected:
\be
  \frac{d\lambda}{d\zeta} = -\lambda^2 + \lambda_g^2, \qquad
  \lambda_g =\frac1{2\pi\sqrt{g}} ,
\label{Fin}
\ee
%where $\zeta =\ln(1/E\tau_{\rm tr})$.
and $\lambda(\zeta=0)=\lambda_n$. The fixed point solution of
Eq.~(\ref{Fin}), $\lambda=\lambda_g$, is too small to fulfil both
the above inequalities together. Therefore, typically the
approximation of single-parameter RG for $G_A$ is not valid, and we
should reconsider the problem of the subgap N-S conductance in the
presence of three different effects acting simultaneously:
%(i) disorder-enhanced multiple Andreev reflections~\cite{nazarov1} which
%increase $G_A(\omega)$ when the coherence length $\sqrt{D/\omega}$
%grows; (ii) Cooper-channel repulsion $\lambda$ which reduces
%$G_A$~\cite{SpiNazSto,FL,Qproxy}; (iii) quantum fluctuations of the
%phase $\theta(\tau)$ which destroy coherence between Andreev
%reflections and suppress $G_A(\zeta)$ at long time scales.
\begin{itemize}
\item[(i)]
disorder-enhanced multiple Andreev reflections~\cite{nazarov1} which
increase $G_A(\omega)$ when the coherence length $\sqrt{D/\omega}$
grows;
\item[(ii)]
Cooper-channel repulsion $\lambda$ which reduces
$G_A$~\cite{SpiNazSto,FL,Qproxy};
\item[(iii)]
quantum fluctuations of the
phase $\theta(\tau)$ which destroy coherence between Andreev
reflections and suppress $G_A(\zeta)$ at long time scales.
\end{itemize}
To treat all these effects together, we employ the functional RG method for
the proximity-effect action in the Keldysh form~\cite{Qproxy}.

Like in the simplified model~\cite{FL} discussed above, the constant
${\cal C}(0)$
is determined (with exponential accuracy) by the value of time $t^*$
%% such that $G_A(\zeta=\ln\omega_d t^*)$ decreases down to values $\leq 1$,
when $G_A(\zeta=\ln\omega_d t^*)$ becomes of the order of
1, since at longer times $C_0(t)$ decays fast. However, the equation
for $G_A(\zeta)$ is much more complicated now as it includes an
infinite set of parameters. To derive the corresponding RG equations,
we start from the Keldysh action for a SC island in contact with a
disordered metal, derived in Ref.~\cite{FLS}. It can be represented
as a sum $S =S_{\rm bulk} + S_{\rm bound}$ of the bulk and boundary
[the last term in Eq.~(\ref{Keld_all})] contributions:
\bea
S = \frac{i\pi\nu}{4}
\Tr \left[
D (\nabla Q)^2
+ 4i \bigl(
i\tau_z \partial_t + \tensor{\phi} + \tensor{\Delta} \bigr) Q
\right]
\nonumber \\
+ \Tr \roarrow{\phi}^T V^{-1} \roarrow{\phi} + \frac{2\nu}{\lambda}
\Tr \roarrow{\Delta}^+ \sigma_x \roarrow{\Delta} - \frac{i\pi G_T}{4}
\Tr Q_S Q .
\label{Keld_all}
\eea
The bulk action, $S_{\rm bulk}$, is a functional of three fluctuating
fields: the matter field $Q(\br,t,t')$ in the film (its average value
gives the time-domain representation of the electron
Green function $G({\bf r,r'})$ at ${\bf r}= {\bf r'}$), the
electromagnetic potential $\roarrow\phi(\br,t)$, and the order
parameter field $\roarrow\Delta(\br,t)$ used to decouple the quartic
interaction vertex in the Cooper channel.
$Q(\br,t,t')$ is a matrix in the
time domain, and in the direct
$4\times4$-dimensional
product $K\otimes N$ of the Keldysh and Nambu-Gor'kov spaces. Pauli
matrices in the $K$ and $N$ spaces are denoted by $\sigma_i$ and
$\tau_i$, respectively.
The field $Q$ satisfies a nonlinear constraint $Q^2=1$ and can be
parametrized as $Q=e^{-W/2}\Lambda e^{W/2}$ with $\{W,\Lambda\}=0$,
where $\Lambda = \Lambda_0 \tau_z$ is the metallic saddle point and
\be
  \Lambda_0(\eps) =
  \left( \begin{array}{cc} 1 & 2F(\eps) \\ 0 & -1 \end{array} \right)_K .
\label{Lambda0}
\ee
%$\Lambda_0(\eps) = \tau_0\sigma_z + 2\sigma_+ F(\eps)$,
The matrix $F(\eps) = \tau_0 f(\eps) +\tau_z f_1(\eps) $ has
the meaning of a generalized distribution function. The object
$\roarrow\phi=(\phi_1, \phi_2)^T$ is a vector in the Keldysh space,
with $\phi_1, \phi_2$ being the classical and quantum components of
the $\phi$-field. $\tensor\phi$ is a shorthand notation for the
matrix $\tensor\phi = \phi_1 \sigma_0 + \phi_2 \sigma_x$. Similarly,
$\roarrow\Delta=(\Delta_1, \Delta_2)^T$, and $\tensor\Delta$ stands
for a $4\times4$ matrix $\tensor\Delta = [\tau_+\Delta_1 -
\tau_-\Delta_1^*]\sigma_0 + [\tau_+\Delta_2 -
\tau_-\Delta_2^*]\sigma_x$, where $\tau_\pm \equiv (\tau_x \pm
i\tau_y)/2$. In terms of the $\sigma$-model action (\ref{Keld_all}),
diffuson and Cooperon collective modes
of the electron system are described as slow fluctuations of the
$Q$-matrix over the manifold $Q^2=1$. The last
(boundary)
term in Eq.~(\ref{Keld_all})
describes an elementary tunneling process between the SC island and
the N metal. The matrix $Q_S$ describes the state of the SC island.
At the low-energy scales $\eps \ll |\Delta_{SC}|$ it is expressed via
the phase $\theta(t)$:
\be
  Q_S = - i \tau_+ e^{i\tensor\theta} + i \tau_- e^{-i\tensor\theta},
\label{QS}
\ee
%$Q_S = - i \tau_+ e^{i\tensor\theta} + i \tau_- e^{-i\tensor\theta}$,
where $\tensor\theta = \theta_1\sigma_0
+ \theta_2\sigma_x$.

The action (\ref{Keld_all}) contains a fluctuating scalar potential
field $\roarrow\phi$ accounting the direct Coulomb interaction in the
density-density channel.
Major effects of this interaction are: (i) local electroneutrality of
electron liquid at low frequencies, and (ii) zero-bias anomaly in the
tunneling DOS~\cite{AAL}.
Both effects can be taken care of by means of a special gauge
transformation~\cite{KA}: $Q_{tt'} \to e^{i\tensor K(t)\tau_z}
Q_{tt'}e^{-i\tensor K(t')\tau_z}$, and $\roarrow\phi(t) \to
\roarrow\phi(t) + \partial_t\roarrow K(t)$. The ``Coulomb phase"
$\roarrow K(t)$ is a linear functional of $\roarrow\phi(t)$ chosen
in a way to cancel all terms which are linear both in $\roarrow\phi
(t)$ and in the matrix field $W({\bf r},t,t')$. Then the effect (i)
is contained in the tree level of the transformed effective action,
whereas (ii) comes from the simplest loop correction~\cite{KA,FLS}.
After the above gauge transformation the phase $\roarrow\theta_j(t)$
of the $j$-th island
enters the action in the combination $\roarrow\theta_j(t) -2\roarrow
K(t,{\bf r}_j)$ only. Now the important point of our discussion comes about:
the phases $\roarrow\theta_j(t)$ of each island are
 not fixed by any external source, and should be integrated out.
Thus the shift of integration variable
$\roarrow\theta_j(t) \to \roarrow\theta_j(t) -2\roarrow K(t,{\bf r}_j)$ eliminates
$\roarrow K(t,{\bf r}_j)$ from the action, together with both effects (i) and
(ii). In other terms, the present problem of unconstrained phase
$\roarrow\theta(t)$ fluctuations can be treated as if it would be no
Coulomb interaction, since Gaussian terms in the action containing
electric field are decoupled from the redefined $\roarrow\theta(t)$
variable. It is thus legitimate to neglect electroneutrality and
calculate frequency-dependent subgap conductance $G_A(\omega)$ as if
the outer normal contact would be placed at the distance $R_\omega =
\sqrt{D/\omega}$ from the SC island.
We emphasize that the same would be wrong for a usual problem of N-S
conductance between contacts with fixed voltages, where the full size
of the N film, $L \gg R_\omega$, does enter the result, adding the
term $(R_\Box/2\pi)\ln(L/R_\omega)$ into the resistance,
cf.~Sec.~VI of Ref.~\cite{Qproxy}.

Next we use the RG method to integrate consecutively over fast
degrees of freedom in the action (\ref{Keld_all}), which is defined
with $\omega_d$ being a high-energy cutoff. At each step of the RG
procedure one has to eliminate fast modes $W_{\eps_1\eps_2}$ in the N film,
with either $\Omega^* > \max(Dq^2,\eps_1-\eps_2)>\Omega$ (for diffusons) or
$\Omega^* > \max(Dq^2,\eps_1+\eps_2)>\Omega$ (for Cooperons), and
fast (with $\Omega^* > \omega > \Omega$) fluctuations of the
order-parameter phase $\roarrow\theta$ on the SC island (where
$\Omega$ is the running infrared RG cutoff). The above integration
results in a correction to the action of slow variables proportional
to the increment of the ``logarithmic time" $\Delta\zeta =
\ln(\Omega^*/\Omega)$.
The structure of the boundary term in the action (\ref{Keld_all}) is
not reproduced under the RG~\cite{FLS}, instead higher-order terms
$\Tr(Q_SQ)^n$ are generated, which are all relevant in the case of
strong proximity effect. The full boundary action can be written in
the form~\cite{Qproxy}
\be
  S_{\rm bound} = \sum_{n=1}^\infty S_n
  = -i \pi^2 g \sum_{n=1}^\infty \gamma_n(\zeta) \Tr (Q_SQ)^n .
\label{sg}
\ee
At the energy scale $\omega_d$, the multicharge action (\ref{sg})
reduces to the last term in Eq.~(\ref{Keld_all}), that is,
$\gamma_1(0)=G_T/4\pi g$ and $\gamma_{n\geq2}(0)=0$.
Integration over all fast modes down to the energy scale
$\omega = \omega_d e^{-\zeta}$ defines the action (\ref{sg}) with the
set of parameters $\gamma_n(\zeta)$.

In the model~\cite{FL} of large $\lambda_n$ discussed above,
separation of scales was possible:
at relatively short time scales
the term $S_2 \propto \gamma_2\Tr{(Q_SQ)^2}$
% it was sufficient to integrate over Cooperon modes only.
% Under the RG the term $\gamma_2\Tr{(Q_SQ)^2}$
was generated under the action of the RG~\cite{FLS},
that lead to a constant value of $G_A \gg 1$
(other $\gamma_{n\geq 3}$ were still small).
At longer time scales fluctuations of
$\theta$ became important, being determined by the
action~(\ref{Action1}).
In the full problem considered now, all parameters $\gamma_n$ are
important, and all types of fluctuations should be considered
simultaneously. The corresponding RG equations for the ``charges''
$\gamma_n(\zeta)$ had been derived in Ref.~\cite{Qproxy} for the case when
the SC island is connected to an external circuit and its phase is fixed by
an external bias $V$ through the Josephson relation $d\theta/dt=2eV$.
In the absence of an external contact, all phases do fluctuate
and these fluctuations lead to an additional logarithmic renormalization
of the action.

\begin{figure} 
\refstepcounter{figure} \label{F:ph-diag1}
\centerline{\epsfbox{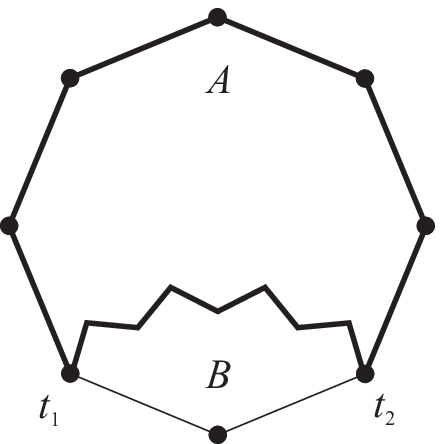}}
\vspace{2mm}
\small FIG.\ \arabic{figure}.
An example of averaging $\langle S_n\rangle$ over phase fluctuations: $n=8$, $p=5$
(see text for details).
Dots denote $Q_S(t)$, straight lines denote $Q(t,t')$, and the zigzag 
line stands for the correlator (\protect\ref{theta-corr}) of fast phases
taken from $Q_S(t_1)$ and $Q_S(t_2)$. 
The fast energy $\omega$ runs over the zigzag line and the part $A$ 
of the diagram, the corresponding elements being shown by thick lines.
\end{figure}

To calculate the corresponding correction one has to average the action
(\ref{sg}) over high-frequency fluctuations of the phase $\roarrow\theta$.
Its propagator is determined by the Gaussian part of the same action
and has the form:
\be
  \langle \theta_i(\omega) \theta_j(-\omega) \rangle =
  \frac{2\Pi_\omega^{ij}}{\omega G_A(\zeta_\omega)},
\label{theta-corr}
\ee
where $G_A(\zeta_\omega)=
16\pi g\sum_{n=1}^\infty (-1)^n n^2 \gamma_{2n}(\zeta_\omega)$
is the scale-dependent Andreev conductance,
$\zeta_\omega=\ln(\omega_d/\omega)$,
and
\be
  \Pi_\omega =
  \left( \begin{array}{cc}
    2\coth(\omega/2T) & 1 \\
    -1 & 0
  \end{array} \right) .
\ee
Let us consider the $n$-th term of the action (\ref{sg}), 
$S_n = -i \pi^2 g \gamma_n \Tr (Q_SQ)^n$,
and find the one-loop correction from it, $\Delta^{(n)}\gamma_k$,
to the parameters $\gamma_k$ due to phase fluctuations with frequencies
$\omega$ in the range $\Omega<|\omega|<\Omega_*$. 
This correction will be proportional
to $\Delta\zeta = \ln(\Omega_*/\Omega)$.
To find it, one has to separate $\tensor{\theta}$
into the sum of the slow
part $\tensor{\tilde\theta}$ (with frequencies $|\omega|<\Omega$),
and the fast part $\tensor{\theta'}$ (with $\Omega<|\omega|<\Omega_*$),
expand $S_n$ to the second order in $\tensor{\theta'}$ and average over it
with the correlator (\ref{theta-corr}), taking into account all possible
pairings. 
There are two types of pairing: two fast variables $\tensor{\theta'}$
can belong to the same matrix $Q_S$ [cf.\ Eq.~(\ref{QS})],
or to two different $Q_S$.
The first case is simple: on averaging over $\tensor{\theta'}$ each of the
terms $S_n$ reproduces its structure, and the whole effect of fluctuations
is to modify $\gamma_n$:
$\Delta^{(n)}\gamma_{n} = - 2n\gamma_n\Delta\zeta/\pi G_A(\zeta)$.
The pairing of the second type involves averaging over the
fast variables $\tensor{\theta'}(t_1)$  and $\tensor{\theta'}(t_2)$
which belong to  different $Q_S$.  Therefore
these two $Q_S$ subdivide the whole cyclic product $(Q_SQ)^n$ under
the trace into two parts:
$\Tr(Q_SQ)^n = \Tr Q_S(t_1) A(t_1,t_2) Q_S(t_2) B(t_2,t_1)$,
where $A=Q(Q_SQ)^p$, $B=Q(Q_SQ)^{n-p-2}$, 
and $p$ is an integer number which distinguishes different pairings.
An example with $n=8$ and $p=5$ is shown in Fig.~\ref{F:ph-diag1}. 
For certainty we will assume that the fast energy $\omega$
runs over the matrix $A$ (the opposite situation can be treated
analogously by replacing $p\leftrightarrow n-p-2$).
Within the RG precision, fast bulk fields $Q$ entering the product $A$
can be substituted by $\Lambda(\omega)=\Lambda_0(\omega) \tau_z$
(we will explain this below) while slow matrices $Q_S$ entering $A$
can be taken at coinciding times $t\approx t_1\approx t_2$.

Averaging is performed with the help of Eq.~(\ref{QS})
and the following algebraic relations 
(which can be proved by direct calculation):
\bea
&& \Bigl\langle
    \tensor{\theta'}
    \cdot
    e^{\pm i\tensor{\tilde{\theta}}} \Lambda_0
    \Bigl(
      e^{\mp i\tensor{\tilde{\theta}}} \Lambda_0
      e^{\pm i\tensor{\tilde{\theta}}} \Lambda_0
    \Bigr)^k
    e^{\mp i\tensor{\tilde{\theta}}}
    \cdot
    \tensor{\theta'}
  \Bigr\rangle
  = 0 ,
\label{theta-aux1}
\\
&& \Bigl\langle
    \tensor{\theta'}
    \cdot
    \Bigl(
      e^{\pm i\tensor{\tilde{\theta}}} \Lambda_0
      e^{\mp i\tensor{\tilde{\theta}}} \Lambda_0
    \Bigr)^k
    e^{\pm i\tensor{\tilde{\theta}}}
    \cdot
    \tensor{\theta'}
  \Bigr\rangle
  = \frac{4 \Delta\zeta}{\pi G_A} e^{\pm i\tensor{\tilde{\theta}}} .
\label{theta-aux2}
\eea
According to these equations, only pairings with odd $p=2k-1$
produce a nonzero contribution.
For them we have
\be
  {\cal N}_k \equiv \corr{Q_S \Lambda (Q_S \Lambda)^{2k-1} Q_S}
  = (-1)^{k+1} \frac{4 \Delta\zeta}{\pi G_A} Q_S ,
\label{QAQ}
\ee
where fast $\tensor{\theta'}$ are taken from the first and the last $Q_S$
under the correlator, and the sign $(-1)^k$ results from commutation
of Pauli matrices in the Nambu space.
Then, collecting all combinatorial factors, one gets
$\Delta S = (-1)^k in\gamma_n (4\pi g\Delta\zeta/G_A) \Tr (Q_SQ)^{n-2k}$.
We see that the paring with $p=2k-1$ modifies $S_{n-2k}$ with
$\Delta^{(n)}\gamma_{n-2k} = (-1)^{k+1} 4n\gamma_n\Delta\zeta/\pi G_A(\zeta)$.
Thus, under the action of the RG, the term $S_n$ generates
the descending series of terms: $S_n$, $S_{n-2}$, $S_{n-4}$, \dots{}
The RG equation for the coefficient $\gamma_n$ reads
\be
  \Delta\gamma_{n} = -\frac{2\Delta\zeta}{\pi G_A(\zeta)}
  \left(n\gamma_{n} + 2\sum_{k=1}^\infty (-1)^k
  (n+2k)\gamma_{n+2k}\right) .
\label{mrg}
\ee

Finally, we have to explain the validity of the substitution
$Q\mapsto\Lambda$ in the ``fast'' product $A$. Such reduction implies
that no slow diffusons or Cooperons could be connected to the
part $A$ of the diagram on further steps of the RG.
This statement is evident for Cooperons, but not for diffusons.
Indeed, Cooperons depend on the fast frequency $\eps_1+\eps_2=2\omega$,
whereas diffusons depend on $\eps_1-\eps_2$ which can be slow, and thus
might contribute to the RG equation on further steps.
Actually, however, they do not. 
The situation here is very similar to the one
described during the derivation of the $\lambda$ term in the RG,
cf.\ Sec.~V of Ref.~\cite{Qproxy}: 
each diagram containing  a diffuson in the ``fast'' product $A$ produces 
several nonzero contributions, but their sum vanishes. 

\begin{figure} 
\refstepcounter{figure} \label{F:ph-diag2}
\centerline{\epsfbox{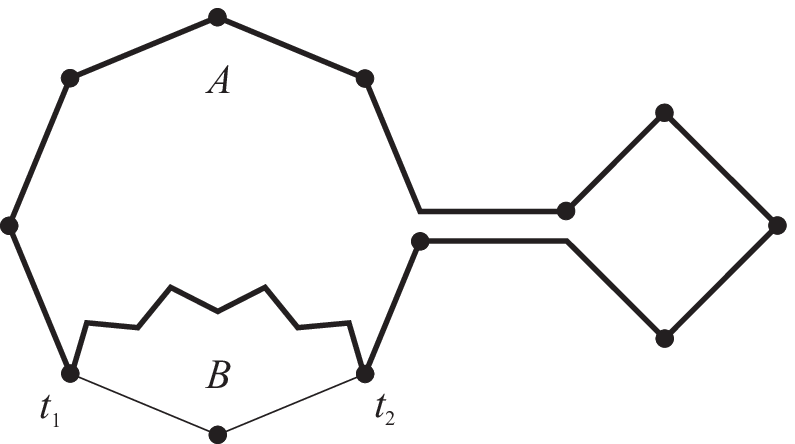}}
\vspace{2mm}
\small FIG.\ \arabic{figure}.
An example of diffuson pairing of a matrix $Q$ from the fast part $A$
of the diagram shown in Fig.~\protect\ref{F:ph-diag1} to the object
$\Tr(Q_S\Lambda)^4W$. The double line denotes the diffuson propagator.
\end{figure}

To prove this cancellation property, it is sufficient to 
consider the case when one $Q$ from the part $A$ is connected 
to some other object $O$ by a diffuson, cf.\ Fig.~\ref{F:ph-diag2}.
Since the pairing between $S_n$ and the object $O$ is of diffusonic
character, the latter ought to have the form 
$O \propto \Tr (Q_S\Lambda)^{2m}W$
with even number of $Q_S$. Pairing is performed with the help of the 
contraction rules from Ref.~\cite{Qproxy} and yields two terms:
\be
  \Tr Q_S(t_1) B Q_S(t_2) 
  \left[ 
    (\Lambda Q_S)^{2k+2m-1}
    -
    (\Lambda Q_S)^{2k-2m-1}
  \right] \Lambda .
\label{-}
\ee
To obtain the final result, 
the expression (\ref{-}) should be averaged over phase fluctuations. 
Note that Eq.~(\ref{QAQ}) is valid for any integer $k$. 
Therefore, averaging (\ref{-}) with the help 
of Eq.~(\ref{QAQ}) yields ${\cal N}_{k+m}-{\cal N}_{k-m}=0$, q.~e.~d.

To simplify the RG equation, we introduce, following Ref.~\cite{Qproxy}, 
%According to Ref.~\cite{Qproxy}, it is convenient to introduce 
a function
of an auxiliary continuous variable $x$ according to the definition
$u(x,\zeta) = \sum_{n=1}^\infty n \gamma_n(\zeta) \sin nx$. 
Then the full RG equation for the function $u(x,\zeta)$ reads:
\be
  u_\zeta + u u_x + \lambda(\zeta)\, u(\pp,\zeta) \sin x =
  - \frac{2}{\pi G_A(\zeta)}{\cal F}[u(x,\zeta)] ,
\label{urgl}
\ee
where ${\cal F}[u(x,\zeta)] \equiv [u(x,\zeta)\tan x-u(\pp,\zeta)\sec x]_x$,
and the initial condition is $u(x,0) = (G_T/4\pi g)\sin x$.
The functional RG equation with the
 L.H.S. of Eq.(\ref{urgl}) and zero in the R.H.S. was derived in~\cite{Qproxy}.
 In the present problem the R.H.S. describes the effect of phase
 fluctuations; it
is  obtained as the Fourier-transform of (\ref{mrg}).
The scale-dependent subgap conductance $G_A(\zeta)$ is determined by the
solution of Eq.~(\ref{urgl}) as
$ G_A(\zeta) = 4\pi g u_x(\pp,\zeta)$.

To find the parameter ${\cal C}(0)$ with exponential accuracy, we
integrate Eq.~(\ref{urgl}) together with Eq.~(\ref{Fin}) for
$\lambda(\zeta)$. Written in the rescaled variables $s =
\zeta/2\pi\sqrt{g}$, $w(x,s) = 2\pi\sqrt{g} u(x,\zeta)$, and
$\tilde\lambda = \lambda/\lambda_g$, Eq.~(\ref{urgl}) acquires the
form
\be
  w_s + w w_x =
  - 2\frac{{\cal F}[w(x,\zeta)]}{w_x(\pp,s)} - \tilde\lambda(s)
  w(\pp,s)\sin{x} ,
\label{rgw}
\ee
with the initial condition $w(x,0) = A\sin{x}$, where $A\equiv
G_T/2\sqrt{g} \gg 1$.
The solution of Eq.~(\ref{rgw}) weakly depends on the ratio
$\tilde\lambda(0) = \lambda_n/\lambda_g$ which is assumed to be not
very large. At $s \ll 1$, the function $w(x,s)$ is close to the
solution of Eq.~(\ref{rgw}) with zero R.H.S., which, at $s \geq
A^{-1}$, is given by $w(x,s) \approx x/s$ for $x\in(0,\pi)$. As $s$
grows, the R.H.S. terms become increasingly important, and eventually reduce
$G_A(\zeta) = 2\sqrt{g}w_x(\pp,s) $ down to the value of the order of
1 at the critical value of $\zeta^* = 2\pi\sqrt{g} s_c$.
The value of $s_c \sim 1$ was determined, for several values of
$\tilde\lambda(0)$, via numerical solution of Eq.~(\ref{rgw}) in the
limit $A\to\infty$:
\be
  s_c = \cases{
    1.25, & for $\tilde\lambda(0) = 0$; \cr
    1.17, & for $\tilde\lambda(0) = 1$; \cr
    1.13, & for $\tilde\lambda(0) = 2$.
  }
\label{scc}
\ee
As it was explained above, the value of $\zeta^* = 2\pi\sqrt{g} s_c$ determines
${\cal C}(0)$ with exponential accuracy:
\be
{\cal C}(0) =  \frac{{\cal B}}{\omega_d }e^{2\pi \sqrt{g} s_c} ,
\label{C0res}
\ee
where pre-exponential factor $\cal B$ is left undetermined 
(to find it, the two-loop RG calculation would be necessary).
Combining  Eqs.~(\ref{transition}), (\ref{J}) and (\ref{C0res}),
we find that macroscopic superconductivity in the array is destroyed
when the distance $b$  between neighbouring islands becomes
larger than the critical value $b_c(g)$ defined as
\be
  \frac{b_c(g)}{d} \approx
%  \frac{\pi^2}{2}\frac{\sqrt{{\cal B}}}{(\pi\sqrt{g} s_c)^{1/2}}
  \frac{\pi^2}{2} \left( \frac{{\cal B}\sqrt{g}}{\pi s_c} \right)^{1/2}
  e^{\pi \sqrt{g} s_c} .
\label{bcg}
\ee
Finally, inverting relation (\ref{bcg}) one finds the
 result (\ref{main}) for the critical film conductance $g_c(b)$,
 with
\be
  {\cal G}_c = \frac1{s_c^2} =
  \cases{
    0.64, & for $\tilde\lambda(0) = 0$; \cr
    0.73, & for $\tilde\lambda(0) = 1$; \cr
    0.79, & for $\tilde\lambda(0) = 2$.
  }
\label{G}
\ee
and $\tilde{d} = d \cdot \pi^{3/2}{\cal B}^{1/2}g^{1/4}/2s_c^{1/2}$.

At $g > g_c(b)$ macroscopic superconductive transition occurs at $T >
0$. Close to the critical point (\ref{main}), at $b \leq b_c(g)$, the
transition temperature is primarily determined by the temperature
dependence of ${\cal J}(T)$ which is reduced compared to ${\cal J}(0)$
since couplings $J(r)$ are strongly suppressed at $r \geq L_T$. At low
temperatures
\be
  {\cal J}(T) = {\cal J}(0)\frac{\ln(L_T/b)}{\ln(L_T/d)}.
\label{JT}
\ee
The same expression, with $L_T$ replaced by $l_H = \sqrt{\pi\hbar c/e H}$,
determines ${\cal J}(0,H)$ in the presence of a weak transverse magnetic field
$H$. To find the critical temperature $T_c(b)$ and the $T=0$ value of
the upper critical field $H_{c2}$, one uses
Eq.~(\ref{transition}) together with Eqs.~(\ref{C0res}), (\ref{JT}).
The result is that both
$T_c(g)$ and $H_{c2}(g)$ scale in the same way, and drop fast at
$b \to b_c(g)$:
\be
  \ln\frac{T^*}{T_c} \approx \ln\frac{\Phi_0}{H_{c2}b^2} \approx
  \frac{2\ln(b/d)}{b_c^2(g)/b^2-1} ,
\label{THg}
\ee
where $T^* = \hbar D/b^2$, and $\Phi_0=hc/2e$ is the flux quantum.
The form of Eqs.~(\ref{THg}) suggests that
the behavior of the array near the quantum critical point $b = b_c(g)$
can be understood in terms of the standard BCS-type theory with a
{\it macroscopic} effective Cooper attraction $\lambda_{\rm eff}$,
 which vanishes at $b \to b_c(g)$. 
Eq.~(\ref{THg}) is valid for $b/b_c(g) \geq
\left(2\ln(b/d)\right)^{-1/2}$. This inequality ensures that $T_c$ is
small compared both to $ T^*$ (under this condition the proximity
coupling is long-range) and to $\hbar/{\cal C}(0)$. The latter
condition allows one to approximate ${\cal C}(T)$ by ${\cal C}(0)$
while deriving Eq.~(\ref{THg}).
At shorter $b \ll b_c(g)\sqrt{\lambda_g}$, the transition occurs at
$T_c \sim T^*$. Here $L_{T_c} \sim b$, the MFA is not applicable and
the transition is governed by thermal fluctuations.
Similarly, the characteristic scale of magnetic field
which affects superconductive state in such an array is just $\Phi_0/b^2$:
at higher fields formation of a superconductive glass state is expected.
Additional limitation for our results from the higher-$T$ side is due to
the neglect of quasiparticles inside SC islands, the corresponding temperature
scale being $T_{\rm parity} = \Delta_{SC}/\ln(\nu V_i\Delta_{SC})$.
On the other hand, the conclusion about metallic nature of the system's
state at $ b > b_c(g)$ is limited to the temperature scale
$T \geq T_{\rm loc} \sim \hbar\omega_d e^{-\pi^2g}$,
since we neglected weak localization effects (the value of $g$ entering this
estimate is defined at the length scale of the order of $d$).
Determination of weak-localization corrections to the conductance of an
array with $b > b_c(g)$ would need the account of dephasing effects
due to fluctuations of island's phases.
This is an interesting problem which we left for the future studies.

Finally, we comment briefly on the similar problem of small superconductive 
grains of radius $d$ immersed into a 3D metal with bulk resistivity $\rho$.
In this case the simple method of Ref.~\cite{FL}
can be used. The characteristic Coulomb energy $\hbar/{\cal C}(0) \approx
\hbar\omega_d \exp(- \pi\hbar/4e^2R_A)$ in the 
tunneling limit $R_T \gg R_N$; here $R_A =R_T^2/R_N$ and $R_N = \rho/4\pi d$.
Proximity coupling energy is  estimated as 
$J(r) \sim (\hbar/e^2R_T)^2/r^3\nu$. The  point of the SC-M transition
is then given by $\hbar\rho/16e^2d R_T^2 \cong 3\ln{b/d}$. 

In conclusion, we developed a theory of quantum superconductive-metal
transition in a 2D proximity-coupled array. The critical resistance
$R_{\Box c}$
is non-universal and small compared to the quantum resistance. Near
the quantum critical point the system behaves as a BCS-like
superconductor with the effective Cooper attraction constant
vanishing at $R_{\Box} \to R_{\Box c}$.

We are grateful to A.~Kamenev and Yu.~V.~Nazarov for useful
discussions. This research was supported by the NSF grant DMR-9812340
(A.~I.~L.), RFBR grant 98-02-16252, NWO-Russia collaboration grant,
Swiss NSF-Rus\-sia collaboration grant 7SUPJ062253.00,
and by the Russian Ministry of Science via the project
``Mesoscopic electron systems for quantum computing"
(M.~V.~F. and M.~A.~S.).

%\end{thebibliography}

\end{multicols}

\end{document}